# Smartphone Sensing Platform for Emergency Management

**Jaziar Radianti**
CIEM, University of Agder, Norway
jaziar.radianti@uia.no

**Julie Dugdale**
Grenoble 2 University/Grenoble Informatics Laboratory (LIG), France
julie.dugdale@imag.fr

**Jose. J. Gonzalez**
CIEM, University of Agder, Norway
jose.j.gozalez@uia.no

**Ole-Christoffer Granmo**
CIEM, University of Agder, Norway
ole.granmo@uia.no

**ABSTRACT**

The increasingly sophisticated sensors supported by modern smartphones open up novel research opportunities, such as mobile phone sensing. One of the most challenging of these research areas is context-aware and activity recognition. The SmartRescue project takes advantage of smartphone sensing, processing and communication capabilities to monitor hazards and track people in a disaster. The goal is to help crisis managers and members of the public in early hazard detection, prediction, and in devising risk-minimizing evacuation plans when disaster strikes.

In this paper we suggest a novel smartphone-based communication framework. It uses specific machine learning techniques that intelligently process sensor readings into useful information for the crisis responders. Core to the framework is a content-based publish-subscribe mechanism that allows flexible sharing of sensor data and computation results. We also evaluate a preliminary implementation of the platform, involving a smartphone app that reads and shares mobile phone sensor data for activity recognition.

**Keywords**

Publish-subscribe; emergency management; mobile sensing; human-centered computing; intelligent systems; sensors; hazard tracking, human tracking,

**INTRODUCTION**

Recently, exploiting smartphone sensors for collecting information has attracted significant attention within the mobile phone sensing and intelligent systems research area. Many possible uses of embedded sensors in smartphones have been extensively explored, study and implemented (Daqing, 2013; Incel, Kose, & Ersoy, 2013; Pessemier, Dooms, & Martens, 2013; Youngki, Younghyun, Chulhong, Jihyun, & Junehwa, 2012). Nevertheless, how to apply smartphone sensors for disaster monitoring, and emergency management, has not yet been widely scrutinized. Owing to the ubiquitous use of smartphones there is an opportunity to introduce ways to employ smartphone sensors to collect information in a crisis.

In a crisis, getting an overview of the hazard situation and ensuring communication between people and stakeholders have been recognized as obstacles hindering a prompt response. Using social media via mobile phones, people have started to organize themselves, to report and indirectly to assist in the crisis management. In the scope of the SmartRescue project our work contributes further to this area by exploiting smartphone technology and developing a smartphone-based platform.

SmartRescue aims at making use of the advanced sensors in newer smartphones to help crisis managers and the public in acute crisis situations. The sensing data is passed through the mobile publish-subscribe (P/S) system, enabling users to publish the sensor readings and others to subscribe to it. The information is used for assessing the hazard and people's location, as well as forming a comprehensive threat picture, with appropriately tailored user-centric evacuation plans. We are using a ship fire as an example for this project. The purpose of this paper is to review existing disaster management platforms, contrasted with SmartRescue, to present the SmartRescue framework, and to discuss the advantages and limitations of the proposed platform.

This paper is structured as follows: Section 2 briefly surveys related work and platforms dedicated for sharing information and dealing with disaster management. Section 3 explains the novel aspects of SmartRescue, presents the core components of the SmartRescue platform, and explains how the platform can help crisis managers/members of the public. Section 4 presents the implementations and the results of the SmartRescue prototyping activities, and provides pointers for further work. The last section summarizes our conclusions.

**RELATED WORK**

In the scope of our work we examined existing emergency management platforms, context aware applications, supporting technologies and the degree of use of mobile phones. iGaDs (Intelligent Guards against Disasters) for example, are used for natural disaster prediction and detection for early disaster alerts over all communication channels. A smart device is deployed to receive, authenticate, and confirm disaster and push alert messages e.g. to the mobile phone users, using a server-to server webhook-based P/S protocol (Ou et al., 2013). The EPIC (Empowering the Public with Information in Crisis) project focuses on providing information supports for the public during a mass emergency. The project promotes the use of social media channels (e.g. Twitter) for mapping, visualizing and monitoring the hazard, which can be done spontaneously by the public such as with Tweak the Tweet (Starbird & Stamberger, 2010), a platform developed under this project.

BRIDGE (Bridging Resources and Agencies in Large-Scale Emergency Management) offers technical and organizational solutions that enhance crisis management in the EU member states, and support multi-agency collaboration in large-scale emergency assistance. UN-SPIDER, a United Nations platform for space-based information for disaster management and emergency response exploits the space-based technologies. Among its goals is to make a platform as a gateway for knowledge acquisition, processing and transfer for the international disaster management community. MRCCFR (Mobile rich media communication and collaboration tool for first responders and a web-based information merging and visualization application) is designed to enable real-time situational awareness by visual means, audio communications, and inter-agency collaboration. It is also intended for tracking people, but not for comprehensive disaster mapping. Users need to tag different symbols for locating the hazard and tracking people (Bakopoulos, Tsekeridou, Giannaka, Tan, & Prasad, 2011). To our knowledge, there is no other platform that is comparable to that developed in the SmartRescue Project, where the backbone is the extensive use of smartphone sensor, intelligent processing and a publish-subscribe system.

**SMARTRESCUE PLATFORM**

The basic of SmartRescue's notion is to use smartphone technology to assist, particularly in the delay phase between the initial time of a crisis, when the victims are left to themselves until the emergency response arrives. The SmartRescue technology maps threats and helps people in the evacuation process. If there are many smartphone users sensing the environment surrounding them, their input can be used as a basis for obtaining a common threat picture, and allow people in the affected area take actions to avoid any hazards such as fire. The project exploits the following trends:

- Embedded sensors in smartphones have quickly developed and increased in numbers. Beyond the traditional audio-visual sensors and the microphone, more sensors have been added to monitor pressure, temperature, light, proximity, barometer, magnetic field, accelerometer, gyroscope and GPS.
- The sensors can be used to map crises through passive and active human-centred sensing, in the spirit of crowd-sourcing, thus providing the potential for exhaustive exploration of the environment, supported by the mobile phones.
- Recent smartphones are powerful computational platforms. They can carry-out complicated tasks without quickly draining the battery or disrupting the smartphone functionality for normal usage.
- Advances in artificial intelligence —many aspects of human intelligence can now be mimicked and applied to many domains. In particular, intelligent processing of data has been brought into the smartphone sensing research area.

All of these trends can be exploited to map crisis situations. The SmartRescue platform is designed for linking the crisis responders to the victims who are close to the hazard. It allows them to communicate the hazard location, shape common threat pictures from the several mobile devices in near real-time through a map, numerical information or graphs, and even develop an up-to-date evacuation plan based on the evolving threat. The platform is meant as a decision support for those who are affected by, or responsible for the disaster:

rescuers and crisis managers, and relies heavily on content-based a publish-subscribe (P/S) system. A P/S system is a communication allowing the users to publish messages or subscribe to message content. In the context of our project the message is mainly sensor data. The Phoenix platform — a P/S middleware implemented in Java, which enables the development of distributed P/S applications, is employed in this work. Phoenix exploits the interaction of three components: *Publishers* that generate information in the form of events, *Subscribers* that subscribe to arbitrary flows of information, and *brokers* that route information from publishers to subscribers.

Fig.1 shows the architecture of the platform. A set of smartphones consist of publishers and subscribers. Under this content-based P/S system, users can publish sensor information through the mobile phone for others, or subscribe to desired sensor information from others. A mobile phone app has been created to carry out P/S tasks of the sensor measurements. For this prototyping purpose, we have only developed the app in an Android platform. The publisher is designed to exploit relevant built-in smartphone sensors and register the external hazardous events, locations, and people's behaviors/movements. The communication occurs via a Broker that forwards the registered sensor measurements to a group of subscribers whenever they match the subscribed keywords. The intended users are the general public (shown in the upper part of figure 1), rescuers and decision makers/crisis managers (the lower part). When a hazard occurs, people close to the hazard can activate the app. Any subscribers connected to the SmartRescue platform via a broker can then be notified.

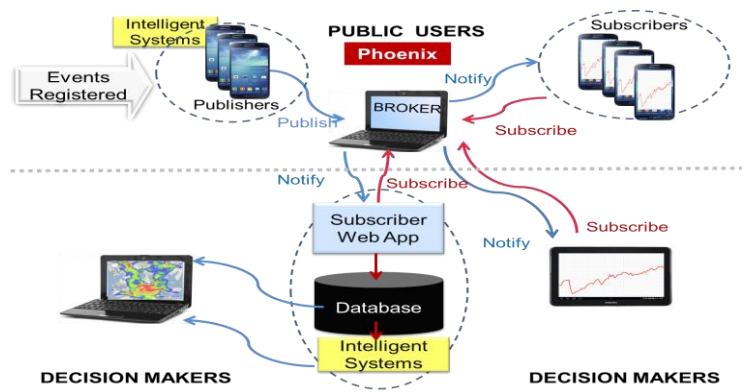

**Figure 1 SmartRescue Platform**

In addition, crisis managers and rescuers have the opportunity to perform the same action as members of the public, i.e. to subscribe to sensor information through mobile devices (tablets, phones), and follow the evolution of the hazard and the movement of people. The platform also allows decision makers to monitor the general picture of the hazard environment and locate people who are proximally close to far away from the hazard's origin.

To support this, a web-based subscriber, which is also connected to the broker, was developed. It has comparable capability to the one deployed in the smartphone, but acts as an intermediary that passes the sensor readings from a group of publishers to the big database environment. It collects, aggregates, and integrates the information, and then presents a summary of useful sensor information. In this way, crisis managers can access, and retrieve information both in its raw format, and as intelligently processed data. The data will be presented and visualized as charts, senders' location on map, and heat maps of a threat to facilitate the observation of the overall crisis situation and people's location. The heat maps have not yet developed.

The platform is designed to be a context aware platform incorporating intelligent systems. The intelligent processing tasks will be distributed in two locations: in the smartphone publishers and in the web subscriber. By locating the intelligent processing in smartphone publishers, it enables mobile subscribers on a real time to receive useful/relevant information only in their user interface. However, the web subscriber that connects to the decision maker interface will still receive the raw data, enabling the decision makers to search for historical information. Concurrently, the intelligent processing placed in the web subscriber aims at shifting some possible excessive computing workloads on smartphone side to optimize the power.

In a crisis, stakeholders, such as ambulance, police and fire department, will typically be present at the hazard location. In addition crisis managers will be collecting data and monitoring the hazard development in order to make decisions. In addition to directly collecting data from smartphones and devising the safest evacuation plan, the SmartRescue platform supports activity recognition via sensors. This can be used for assessing if a person falls and did not move for a period of time, thus possibly indicating a severe injury or death. Decision makers can also quickly be made aware of any possible bottlenecks in the evacuation and an alternative plan can rapidly devised to prevent a worsening of the situation. Note, although the evacuation planning capability has been implemented (Goodwin et. al, 2013; Granmo et al., 2013) we have not yet integrated it into the SmartRescue platform.

**RESULTS AND DISCUSSION**

As previously mentioned, there are two parts to the SmartRescue development: the mobile app part and the web-based part. The development of the SmartRescue app is currently adapted to smartphones running android 4.0 and above; this covers 72% of android phones on the market. We use the Android Developer Tools (ADT), a plugin for Eclipse IDE. For testing our publisher and subscriber apps, we use a Samsung Galaxy S4 GT-I9505, which has the Android 4.2.2 operating system, with Quad-Core processor of capacity 1.9 GHz. One of the main reasons for using this smartphone is that it is equipped with barometer, thermometer and humidity sensors. In addition several Google APIs are employed (i.e. location and sensor APIs, Google Map API v.2, and activity recognition API), which allowed us to access the necessary sensors tested in our project. Regarding the monitoring platform, we have developed a web subscriber that performs the same role as the mobile phone subscriber, which can be seen as a chart and map visualization. Google Map API v.3 is used for the web-based map. The publisher's location, light condition and the user activity can be captured from this web-based map. The additional feature is that the data captured by web subscribers is stored in the database. Currently, we employ Mongo DB for storing the data.

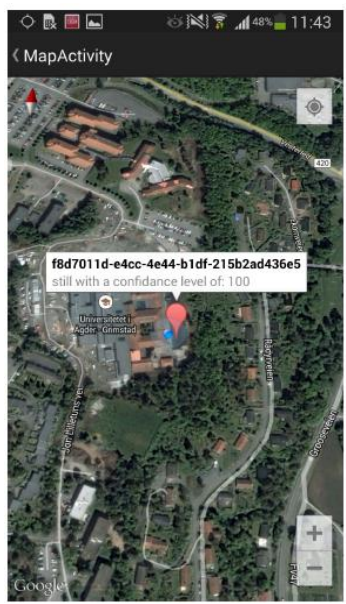

**Figure 2 A screen shot of a subscriber's map**

These APIs are built on top of the previously mentioned Phoenix P/S framework. The Phoenix P/S middleware has been adapted and integrated with the Android app which access the hardware and produce the sensor readings. The publisher and subscriber are integrated in to a single app where the user can decide if they want to set the app as publisher, subscriber or both. The modification of the broker is not needed when we run the infrastructure in a Wi-Fi network. For testing purposes, the broker is installed on an Ubuntu Server 13.04.

The sensor and location information flows from the publisher to the subscriber, and its location can be shown in the map, once the broker identifies the subscriber. At present, we have tested five sensors, i.e. the accelerometer, barometer, thermometer, humidity and light sensors. Figure 2 shows a screenshot of a map in the subscriber app. The blue dot is a standard GPS sign on Google Map that signifies the location of the subscriber. The user can see the location of the subscriber and the information about the person who acts as a publisher. The window that pops up from the publisher's marker is the identity of the publisher and the information about the user's activity (which is stated as "Still with a confidence level of 100"). The user_id is included to enable the subscriber to differentiate multiple publishers. However, to protect the privacy of the user, the app only accesses the installation code, and not more private information inside the phone. In figure 2, the user_id is indicated by the first line of the alphabet/number combination inside the publisher's window.

More sensor information can be obtained by tapping the info window. So far, we display charts, numerical values and text interpretation of the value as a part of the intelligent processes. However, the implementation of the intelligent processes is still simple. The subscriber can assess the user's environment from the measured luminous flux level measured by the light sensor, and by detecting the user's activity (e.g. still, walking, in a car). In general, the platform has performed well in term of the capability to recognize multiple publishers and subscribers, and routing the sensor information. A user can easily setup the broker, publisher, and subscriber as well as obtaining or providing the sensor information, although we notice some delays between the information being published and received. Occasionally, the publisher is still visible for a while on the map of a subscriber when the actual publisher in fact has exited the platform, or there is a span between marker on map and the actual publisher's position. Note the main point with this P/S platform is that, as long as users are connected to the publish-subscribe platform, they do not need to know other people's phone numbers to be able to obtain the information. Likewise, the publishers can just publish the sensor information and leave the routing tasks to the broker who will notify all subscribed smartphones of any relevant information. Hence, as a proof of concept, we consider that the platform is operating as expected, and is a promising base for expanding the SmartRescue functionalities.

The proposed platform is developed with the scalability in mind. In term of sending and receiving the all large sensor data simultaneously with reasonable time interval, the system demonstrates this capability. Moreover, the Phoenix broker can handle many concurrent mobility clients. We tested the system by running the app in a tablet and four smartphones. We also conducted experiments in the machine that runs a publisher and multiple subscribers connected to the broker, due to our limitation with number of our devices for testing. In the machine, we added up to 30 subscribers to the broker, and the system still runs well although some delays can be noticed.

Finally, some technical tasks remain, such as improving the activity and environment recognitions so that we can identify the relevant information for crisis situations, beyond what we can include from activity recognition APIs. In addition, the data aggregation from several smartphone sensors to form the plan common threat pictures is not yet implemented. Incorporating evacuation planning support will also be a part of further improving the SmartRescue platform. To conclude this project we will further address the policy and procedure issue. Although we have been working closely with stakeholders we need to assess in more depth how the platform aligns with current procedure.

**CONCLUSION**

We have developed a publish/subscribe platform for emergency management which, at its heart, is a mix of the use of smartphone sensors, intelligent processing and mapping/visualization. The platform has functioned as expected as a proof of concept. However, some major steps are still required to ensure the usability and user-friendliness of this platform. Therefore, we have involved a user reference group in this project who will provide us with input and feedback so that the platform fits more closely to the user needs.

**ACKNOWLEDGMENTS**

This work was supported in part by the Aust-Agder Utvikling- og Kompetansefond (AAUKF), Norway.